\documentclass[a4paper,12pt]{article}
\pdfoutput=1

\usepackage{undertilde}

\usepackage{amssymb} 
\usepackage{amsmath}
\usepackage{epsfig}
\usepackage{graphicx}

\makeatletter

\@addtoreset{equation}{section}
\makeatother
\def\be{\begin{equation}}
\def\ee{\end{equation}}
\def\bea{\begin{eqnarray}}
\def\eea{\end{eqnarray}}

\def\({\left(}
\def\){\right)}
\def\<{\left<}
\def\>{\right>}

\def\be{\begin{equation}}
\def\ee{\end{equation}}

\def\ben{\begin{eqnarray}}
\def\een{\end{eqnarray}}
\def\({\left(}
\def\){\right)}
\def\<{\left<}
\def\>{\right>}
\def\!{\right|}
\def\|{\left|}

\def\[{\left[}
\def\]{\right]}

\def\+{\bar}
\def\mb{\mathbb}

\def\D{{\cal{D}}}
\def\L{{\cal{L}}}
\def\t{\widetilde}
\def\A{{\cal{A}}}

\def\F{{\mathcal{F}}}

\def\L{{\cal{L}}}

\def\eps{{\cal{\varepsilon}}}

\def\E{{\cal{E}}}

\def\F{{\cal{F}}}

\def\h{\widehat}

\begin{document}

\setlength{\unitlength}{1mm}

\pagestyle{empty}
\vskip-10pt
\vskip-10pt
\hfill 
\begin{center}
\vskip 3truecm
{\Large \bf
Lightlike reduction of the M5 brane}
\vskip 2truecm
{\large \bf
Andreas Gustavsson}
\vspace{1cm} 
\begin{center} 
Physics Department, University of Seoul, Seoul 02504 KOREA
\end{center}
\vskip 0.7truecm
\begin{center}
(\tt agbrev@gmail.com)
\end{center}
\end{center}
\vskip 2truecm
{\abstract We obtain a Lagrangian for a lightlike dimensional reduction of the nonabelian M5 brane theory in six dimension. We assume that the six-manifold has at least one conformal Killing spinor and two commuting lightlike Killing vector fields, and we perform the dimensional reduction along one of these lightlike directions.}

\vfill
\vskip4pt
\eject
\pagestyle{plain}

\section{Introduction}
The lightlike dimensional reduction of the M5 brane results in instanton moduli space dynamics \cite{Lambert:2011gb}. In this paper we will revisit a technical issue regarding the way that selfduality is implemented in these theories. By lightlike reduction, selfduality of the six-dimensional tensor field is inherited as a four-dimensional selfduality condition on the Yang-Mills field strength. There are two independent lightlike directions and the lightlike dimensional reduction results in a five-dimensional theory where one direction is singled out as it originates from the other lightlike direction. We are now facing the problem of how to implement the selfduality equation of motion from a Lagrangian. It turns out that there is a Lagrangian that does precisely this through an auxiliary Lagrange multiplier field $G_{ij}$ that is antiselfdual offshell.

The Lagrangian can not be straightforwardly obtained by dimensional reduction of some Lagrangian for the abelian M5 brane. The reason can be traced to the selfduality of the tensor gauge field in six dimensions, and to the selfduality of the Yang-Mills field in the reduced theory. Let us assume that we have an abelian M5 brane on $\mb{R}^{1,5}$. If we denote the nonselfdual six-dimensional tensor field as $H_{MNP}$, the two lightlike directions as $x^{\pm}$, and define 
\bea
F_{ij} &=& H_{ij-}\cr
G_{ij} &=& H_{ij+}
\eea
then selfduality of $H_{MNP}$ implies selfduality of $F_{ij}$ and antiselfduality of $G_{ij}$. But whereas selfduality of $H_{MNP}$ is an equation of motion in six dimenions, in the dimensionally reduced theory, we will take antiselfduality of $G_{ij}$ as a constraint that holds offshell and selfduality of $F_{ij}$ arises as an equation of motion by varying $G_{ij}$ in the term $\frac{1}{2} G_{ij} F_{ij}$ in the Lagrangian. If dimensional reduction is carried out straightforwardly, then we will get antiselfduality of $G_{ij}$ as an onshell equation of motion. This explains why we need to proceed by some amount of guesswork, if we want to have offshell antiselfduality of $G_{ij}$.  

By making a general ansatz with some arbitary coefficients, we find that for the supersymmetry variation of the fermion we shall have
\bea
\delta \psi &=& - \frac{1}{4} \Gamma_{ij}\Gamma^+ \eps G_{ij} + \frac{1}{4} \Gamma_{ij}\Gamma^- \eps F_{ij} + ....\label{minus}
\eea
The minus sign for the first term is counterintuitive. If we start from the supersymmetry variation in six dimensions  
\bea
\delta \psi &=& \frac{1}{12} \Gamma^{MNP} \eps H_{MNP} + ...
\eea
and decompose this variation into components, then we get 
\bea
\delta \psi &=& \frac{1}{4} \Gamma_{ij} \Gamma^+ \eps G_{ij} + \frac{1}{4} \Gamma_{ij} \Gamma^- \eps F_{ij} + ...
\eea
Instead of reducing the supersymmetry variations, we reduce the selfduality equation of motion 
\bea
H_{MNP} &=& \frac{1}{6} \eps_{MNP}{}^{RST} H_{RST}
\eea
along $x^-$. We then find a Lagrangian that gives those corresponding equations of motion that we want to get. That is, the selfduality equation for $F_{ij}$ (but not for $G_{ij}$) and another equation of motion to be presented below. It turns out that there exists a Lagrangian that gives all the desired equations of motion. It is given by \cite{Lambert:2019jwi}, \cite{Lambert:2020scy}
\bea
\L_A &=& \frac{1}{2} G_{ij} F_{ij} + \frac{1}{2} F_{i+} F_{i+}
\eea
It is important to note the very different roles that these various fields play. While $F_{ij}$ and $F_{i+}$ are usual field strength components in five dimensions, $G_{ij}$ shall be viewed as an auxiliary field that we do not express in terms of some gauge potential. We then supersymmetrize this Lagrangian. This strategy is similar in spirit to what was done in \cite{Linander:2011jy} for a spatial dimensional reduction of the M5 brane on a circle bundle.  It is somewhat similar in spirit to Sen's construction of the selfdual tensor field Lagrangian that also uses an auxiliary offshell selfdual tensor field to implement selfduality on $H_{MNP}$ as an equation of motion \cite{Lambert:2019diy}, \cite{Andriolo:2020ykk}, \cite{Sen:2015nph}, \cite{Sen:2019qit}. We review Sen's construction in the Appendix \ref{Sen}. 
  
In section 2 we present the Lagrangian that we get by lightlike reduction of the M5 brane on $\mb{R}^{1,5}$. We subsequently show that this can be generalized to a nonabelian gauge group. In section 3 we covariantize our flat space result for a rather generic six-manifold with two lightlike Killing vector fields. Here computations are a bit complicated if one chooses an explicit lightcone coordinate system on the six-manifold and of the corresponding five-dimensional base-manifold. Instead we will introduce two lightlike Killing vector fields in the six-manifold and formulate the theory covariantly in six-dimensions. We then impose as a constraint on top of the Lagrangian that all the fields have a vanishing Lie derivative along one of the two lightlike directions. In sections 4 we study reduction of supersymmetry by a lightlike Weyl projection. A detailed computation of the supersymmetry variation of the Lagrangian is presented in Appendix B.

\section{Minkowski space}
We begin with putting the M5 brane on the Minkowski space $\mb{R}^{1,5}$. We assume that the metric is 
\bea
ds^2 &=& - dt^2 + (dx^5)^2 + dx_i dx_i\cr
&=& 2 dx^+ dx^- + dx_i dx_i
\eea
where the two lightlike coordinates are chosen as
\bea
x^{\pm} &=& \frac{1}{\sqrt{2}} \(x^5 \pm t\)
\eea
We define
\bea
F_{ij} &=& H_{ij-}\cr
F_{i+} &=& H_{i+-}\cr
G_{ij} &=& H_{ij+}
\eea
and 
\bea
\eps^{ijkl} &=& \eps^{ijkl+-} 
\eea
Dimensional reduction is performed by putting the derivative along $x^-$ to zero on all fields. The selfduality equation of motion $H_{MNP} = \frac{1}{6} \eps_{MNP}{}^{RST} H_{RST}$ implies the following equations of motion
\bea
F_{ij} &=& \frac{1}{2} \eps_{ijkl} F_{kl}\cr
G_{ij} &=& - \frac{1}{2} \eps_{ijkl} G_{kl}\cr
H_{ijk} &=& - \eps_{ijkl} F_{l+}
\eea
The Bianchi identity $\partial_{[Q} H_{MNP]} = 0$ implies
\bea
\partial_+ F_{ij} + 2 \partial_{[i} F_{j]+} &=& 0\label{Bianchi2a}\\
3\partial_{[i} G_{jk]} - \partial_+ H_{ijk} &=& 0
\eea
These Bianchi identities are of course satisfied offshell, but in addition we will now also assume that the following selfduality equations
\bea
G_{ij} &=& - \frac{1}{2} \eps_{ijkl} G_{kl}\cr
H_{ijk} &=& \eps_{ijkl} F_{l+}
\eea
are satisfied offshell as constraints. Then the remaining equations 
\bea
F_{ij} &=& \frac{1}{2} \eps_{ijkl} F_{kl}\cr
\partial_i G_{ij} &=& \partial_+ F_{j+}\cr
\partial_i F_{i+} &=& 0\label{Gij}
\eea
will be viewed as ordinary equations of motion that we would like to derive from a Lagrangian and Euler-Lagrange equations. Indeed there is such a Lagrangian. It is given by
\bea
\L_A &=& \frac{k}{2} \(G_{ij} F_{ij} + F_{i+} F_{i+}\)\label{LA}
\eea
To this we add the matter part Lagrangian  
\bea
\L_m &=& - \frac{1}{2} (D_i \phi^A)^2 + \frac{i}{2} \bar\psi \Gamma_i D_i \psi + \frac{i}{2} \bar\psi \Gamma_- D_+ \psi\cr
&& + \frac{e}{2} \bar\psi \Gamma_- \Gamma^A [\psi,\phi^A]
\eea
where we assume the gauge group is nonabelian and the covariant derivative is $D_M \Phi = \partial_M - i e [A_M,\Phi]$. The inner product on the gauge Lie algebra, that is the trace, is implicit in our notation. For the supersymmetry variations we make the ansatz
\bea
\delta \phi^A &=& i \bar\eps \Gamma^A \psi\cr
\delta A_i &=& i \bar\eps \Gamma_{i-} \psi\cr
\delta A_+ &=& i \bar\eps \Gamma_{+-} \psi\cr
\delta \psi &=& \frac{a}{4} \Gamma_{ij}\Gamma_- \eps G_{ij} + \frac{b}{4} \Gamma_{ij}\Gamma_+ \eps F_{ij} - c \Gamma_{i}\Gamma_{+-} \eps F_{i+} + \Gamma_i \Gamma^A \eps D_i \phi^A + \Gamma_- \Gamma^A \eps D_+ \phi^A\cr
\delta G_{ij} &=& \frac{i f}{2} \bar\eps \Gamma_k \Gamma_{ij+} D_k \psi + \frac{i f}{2} \bar\eps \Gamma_- \Gamma_{ij+} D_+ \psi\cr
&& + \frac{e}{2} \bar\eps \Gamma_- \Gamma_{ij+} \Gamma^A [\psi,\phi^A]
\eea
There are five scalar fields $\phi^A$ and fermions $\psi$. We use eleven-dimensional gamma matrices satisfying in particular $\{\Gamma^M,\Gamma^A\} = 0$, and the spinors satisfy the eleven-dimensional Majorana condition $\bar\psi = \psi^T C$ and the six-dimensional Weyl projection $\Gamma \psi = \psi$ and $\Gamma \eps = - \eps$ where $\Gamma = \Gamma^{012345}$. When we then make a supersymmetry variation of the Lagrangian, we get the following conditions for the parameters,
\bea
a + k &=& 0\cr
b + kf &=& 0\cr
c - b &=& 0
\eea
from the noncommutator terms, and from the commutator terms we get additional conditions
\bea
e &=& e c\cr
k &=& b\cr
1 &=& k
\eea
From the latter three conditions, we again get $b = c$, which is in agreement with the condition we get in the abelian case, but in addition we get $c = k = 1$ that we do not find in the abelian case. So demanding supersymmetry in the nonabelian generalization gives a more restrictive solution for the coefficients. With $c =k= 1$ the remaining coefficients are
then also uniquely fixed as 
\bea
a &=& -1\cr
b &=& 1\cr
c &=& 1\cr
f &=& -1\cr
k &=& 1
\eea
Within our ansatz, we have found a unique set of parameters for which we have the following nonabelian Lagrangian
\bea
\L &=& \frac{1}{2} \(G_{ij} F_{ij} + F_{i+} F_{i+}\) - \frac{1}{2} (D_i \phi^A)^2 \cr
&& + \frac{i}{2} \bar\psi \Gamma_i D_i \psi + \frac{i}{2} \bar\psi \Gamma_- D_+ \psi + \frac{e}{2} \bar\psi \Gamma_- \Gamma^A [\psi,\phi^A]
\eea
This is invariant under the supersymmetry variations 
\bea
\delta \phi^A &=& i \bar\eps \Gamma^A \psi\cr
\delta A_i &=& i \bar\eps \Gamma_{i-} \psi\cr
\delta A_+ &=& i \bar\eps \Gamma_{+-} \psi\cr
\delta \psi &=& - \frac{1}{4} \Gamma_{ij}\Gamma_- \eps G_{ij} + \frac{1}{4} \Gamma_{ij}\Gamma_+ \eps F_{ij} - \Gamma_{i}\Gamma_{+-} \eps F_{i+} + \Gamma_i \Gamma^A \eps D_i \phi^A + \Gamma_- \Gamma^A \eps D_+ \phi^A\cr
\delta G_{ij} &=& - \frac{i}{2} \bar\eps \Gamma_k \Gamma_{ij+} D_k \psi - \frac{i}{2} \bar\eps \Gamma_- \Gamma_{ij+} D_+ \psi + \frac{e}{2} \bar\eps \Gamma_- \Gamma_{ij+} \Gamma^A [\psi,\phi^A]
\eea

\section{Covariant formulation}
We now turn to a generic Lorentzian six-manifold that we assume has two independent lightlike Killing vectors $u^M$ and $v^M$ respectively. We assume that their inner product
\bea
v_M u^M &=& \lambda
\eea
is everywhere nonvanishing. The two lightlike Killing vectors satisfy the equations,
\bea
v^M v_M &=& 0\\
u^M u_M &=& 0\\
\nabla_M v_N + \nabla_N v_M &=& 0\\
\nabla_M u_N + \nabla_N u_M &=& 0
\eea
Let us now start by the Killing vector equation
\bea
\nabla_M v_N + \nabla_N v_M &=& 0
\eea
If we contract the left-hand side by $u^M$, then we have 
\bea
0 &=& u^M \nabla_M v_N + u^M \nabla_N v_M\cr
&=& u^M \nabla_M v_N + \partial_N \lambda - v^M \nabla_N u_M\cr
&=& u^M \nabla_M v_N + v^M \nabla_M u_N + \partial_N \lambda
\eea
and if we further contract by $u^N$, then the first term vanishes by the Killing equation and the second term vanishes by using $u^N u_N = 0$. We thus find that 
\bea
\L_u \lambda &=& 0\cr
\L_v \lambda &=& 0
\eea
In the Abelian case, we define 
\bea
A_M &=& B_{MN} v^P\label{A1}
\eea
which satisfies $A_M v^M = 0$. Let us now examine the implications of selfduality equation of motion
\bea
H_{MNP} &=& \frac{1}{6} \eps_{MNPRST} H^{RST}
\eea
on the component fields
\bea
F_{MN} &=& H_{MNP} v^P\label{A2}\\
G_{MN} &=& H_{MNP} u^P\label{A3}\\
K_M &=& H_{MNP} u^N v^P\label{A4}
\eea
that we may further expand in traceless and traceparts as
\bea
F_{MN} &=& \t{F}_{MN} + K_M v_N - K_N v_M\cr
G_{MN} &=& \t{G}_{MN} - K_M u_N + K_N u_M
\eea
while $K_M$ is already traceless. Then we may expand 
\bea
H_{MNP} &=& \t{H}_{MNP} + \frac{3}{\lambda} \t{F}_{MN} u_P + \frac{3}{\lambda} \t{G}_{MN} v_P - \frac{6}{\lambda^2} K_M u_N v_P
\eea
The induced selfduality relations are then given by
\bea
\frac{1}{36} \eps_{MNP}{}^{RST} \t{H}_{RST} &=& - \frac{1}{\lambda^2} K_M u_N v_P\cr
\frac{1}{6} \eps_{MNP}{}^{RST} \t{F}_{RS} u_T &=& \t{F}_{MN} u_P\cr
\frac{1}{6} \eps_{MNP}{}^{RST} \t{G}_{RS} v_T &=& \t{G}_{MN} v_P
\eea
If one defines
\bea
\E^{MNRS} &=& \frac{1}{\lambda} \eps^{MNRSTP} v_T u_P
\eea
then one may express these relations as
\bea
\t{H}^{RST} &=& \frac{1}{\lambda^2} \E^{RSTM} K_M\label{t1}\\
\frac{1}{2} \E_{MN}{}^{RS} \t{F}_{RS} &=& \t{F}_{MN}\label{t2}\\
\frac{1}{2} \E_{MN}{}^{RS} \t{G}_{RS} &=& - \t{G}_{MN}\label{t3}
\eea
These definitions (\ref{A1}) and (\ref{A2}) are mutually consistent only if we require that $\L_v F_{MN} = 0$. We impose (\ref{t3}) as well as (\ref{t1}) as an offshell constraints. In addition, we have the usual Bianchi identity $D_{[M} F_{NP]} = 0$ that holds offshell. In Minkowski space we also had the Bianchi identity (\ref{Bianchi2a}) that we should now covariantize as well. To this end, we start by computing the following Lie derivative,
\bea
\L_u F_{MN} &=& \L_v G_{MN} - \nabla_M K_N + \nabla_N K_M + H_{MNP} \L_v u^P 
\eea
To get this result, we have used the Bianchi identity $\nabla_{[Q} H_{MNP]} = 0$. We now see that if $\L_v G_{MN} = 0$ and $\L_v u^M = 0$, then we have the Bianchi identity
\bea
\L_u F_{MN} + 2 \nabla_{[M} K_{N]} &=& 0\label{Bianchi2}
\eea
It is easy to show that $\L_v u^M = 0$ is equivalent to the condition that the Lie derivatives for the two Killing vectors commute, $[\L_u,\L_v] = 0$, which means that we have a torus fibration, with the only peculiar feature being that the vectors that span this torus are lightlike vectors. With these assumptions, we find that the following Lagrangian $\L = \L_A + \L_{CS} + \L_m$, where
\bea
\L_A &=& \frac{1}{2 \lambda} \t{G}^{MN} \t{F}_{MN} + \frac{1}{2\lambda} K^M K_M\cr
\L_{CS} &=& - \frac{1}{4\lambda} \eps^{MNPQRS} \omega(A)_{MNP} \Omega_{QR} u_S\cr
\L_m &=& - \frac{1}{2} (D_M \phi^A)^2 - \frac{R}{10} (\phi^A)^2\cr
&& + \frac{i}{2} \bar\psi \Gamma^M D_M \psi + \frac{e}{2} \bar\psi \Gamma_M \Gamma^A [\psi,\phi^A] v^M\label{Lag}
\eea
is invariant under the following supersymmetry variations
\bea
\delta \phi^A &=& i \bar\eps\Gamma^A\psi\cr
\delta A_M &=& i\bar\eps\Gamma_{MN}\psi v^N\cr
\delta \psi &=&  \Gamma^{MNP}\eps\(- \frac{1}{4\lambda} \t{G}_{MN}v_P +  \frac{1}{4\lambda} \t{F}_{MN} u_P - \frac{1}{\lambda} K_M u_N v_P\)\cr
&& + \Gamma^M \Gamma^A \eps D_M\phi^A - 4 \Gamma^A\eta \phi^A - \frac{i e}{2} \Gamma_M\Gamma^{AB}\eps [\phi^A,\phi^B] v^M\cr
\delta G_{MN} &=& - \frac{i}{2} D_Q \(\bar\eps \Gamma^Q \Gamma_{MNP} \psi\) u^P + \frac{e}{2} \bar\eps\Gamma_Q\Gamma_{MNP}\Gamma^A[\psi,\phi^A] u^P v^Q
\eea
Here we normalize the Chern-Simons three-form such that its variation is given by  $\delta \omega(A)_{MNP} = \delta A_M F_{NP}$ and we define $\Omega_{MN} = \nabla_M\( \frac{1}{\lambda} u_N\) - \nabla_N \(\frac{1}{\lambda} v_N\)$. When we make a supersymmetry variation, we get the result
\bea
\delta \L &=& \frac{i}{2\lambda} \bar\eps \Gamma^{MN} \psi \L_v G_{MN}\cr
&& + 2 e \bar\psi \Gamma^{AB} (\L_v \eps) [\phi^A,\phi^B] + 2 e \bar\psi \Gamma^{AB} \eps [\L_v \phi^A,\phi^B] - e \bar\psi \eps [\L_v \phi^A,\phi^A]\cr
&& + \frac{i}{\lambda} \bar\eps \Gamma^{MN} \psi \L_v \(K_M u_N\)
\label{deltaL}
\eea
Here the Lie derivative of a fermion is defined as $\L_v \eps = v^M \nabla_M \eps + \nabla_M v_N \Gamma^{MN} \eps$. We see that the Lagrangian is invariant if we impose $\L_v = 0$ on all fields, but in addition we need to assume that $\L_v u^M = 0$. We also need to impose that $\L_v \eps = 0$ which is a restriction on the geometry. The details of the computation of the supersymmetry variation of the Lagrangian are presented in Appendix \ref{Lm}.

\section{A lightlike Weyl projection}
We may reduce the amount of supersymmetry by half by imposing the following lightlike Weyl projection
\bea
\Gamma_M \eps v^M &=& 0\label{lightproj}
\eea
Then all the explicit commutator terms will drop out from the supersymmetry variations,
\bea
\delta \phi^A &=& i \bar\eps\Gamma^A\psi\cr
\delta A_M &=& i v_M \bar\eps\psi\cr
\delta \psi &=& \frac{1}{4\lambda} \Gamma^{MNP}\eps \t{F}_{MN} u_P + \Gamma^M \eps K_M + \Gamma^M \Gamma^A \eps D_M\phi^A\cr
\delta G_{MN} &=& - \frac{i}{2} D_Q \(\bar\eps \Gamma^Q \Gamma_{MNP} \psi\) u^P \label{SUSY}
\eea
One may notice that (\ref{lightproj}) implies that
\bea
v^M \nabla_M \eps &=& \frac{1}{4} \nabla_M v_N \Gamma^{MN} \eps
\eea
which can be used to show that when we make a supersymmetry variation of the Lagrangian, one gets
\bea
\delta \L &=& \frac{i}{2\lambda} \bar\eps \Gamma^{MN} \psi \L_v G_{MN} \cr
&& + 2 e \bar\psi \Gamma^{AB} (\L_v \eps) [\phi^A,\phi^B] + 2 e \bar\psi \Gamma^{AB} \eps [\L_v \phi^A,\phi^B] - e \bar\psi \eps [\L_v \phi^A,\phi^A] \cr
&& + \frac{i}{\lambda} \bar\eps \Gamma^{MN} \psi \L_v \(K_M u_N\)
\eea
Although the explicit commutators dropped out from the supersymmetry variations by the lightlike Weyl projection of the supersymmetry parameter, the Lagrangian is of course the same as before, and we can now also see that its supersymmetry variation under the lightlike Weyl projected supersymmetries is also of the same form as before. In particular the Lagrangian can be supersymmetric only if $\L_v \eps = 0$, which is a condition that severly restricts the possible six-manifolds.

\subsection{Further reduction of supersymmetry}
We can avoid having to impose the condition $\L_v \eps = 0$ by further reducing down to the $(1,0)$ tensor multiplet. To this end, we impose a second Weyl projection that breaks the $SO(5)$ R-symmetry down to $SU(2)$ R-symmetry. We define the chirality matrix $\h\Gamma = \h\Gamma^{1234}$ where we put hats on 11d gamma matrices associated with $SO(5)$. We then project the supersymmetry parameter as 
\bea
\h\Gamma \eps &=& - \eps
\eea
We also have the 6d Weyl projection $\Gamma \eps = - \eps$. If we define $\h\Gamma^5$ such that $\Gamma \h\Gamma \h\Gamma^5 = 1$, then we find that $\h\Gamma^5 \eps = \eps$. We also notice that $\h\Gamma^T = C \h\Gamma C^{-1}$ and consequently $\bar\eps \h\Gamma = - \bar\eps$ and $\bar\eps \Gamma^5 = - \bar\eps$. The $(2,0)$ tensor multiplet spinor gets separated into two Weyl components. The component with the same chirality as $\eps$ goes into the vector multiplet spinor that we will continue to denote as $\psi$,
\bea
\h\Gamma\psi &=& - \psi
\eea
We will not consider hypermultiplets. For the vector mulitiplet we have the supersymmetry variations (using the notation $\varphi = \phi^5$ for the fifth scalar field)
\bea
\delta \varphi &=& - i \bar\eps \psi\cr
\delta A_M &=& i v_M \bar\eps \psi\cr
\delta \psi &=& \frac{1}{4\lambda} \Gamma^{MNP}\eps \t{F}_{MN} u_P + \Gamma^M \eps K_M + \Gamma^M \eps D_M \varphi\cr
\delta G_{MN} &=& - \frac{i}{2} D_Q \(\bar\eps \Gamma^Q \Gamma_{MNP} \psi\) u^P
\eea
The vector multiplet Lagrangian is 
\bea
\L &=& \frac{1}{2\lambda} \t{G}^{MN} F_{MN} + \frac{1}{2\lambda} K^M K_M + \L_{CS} - \frac{1}{2} (D_M \varphi)^2 - \frac{R}{10} \varphi^2\cr
&& + \frac{i}{2} \bar\psi \Gamma^M D_M \psi + \frac{e}{2} \bar\psi \Gamma_M [\psi,\varphi] v^M
\eea
We may now introduce a new gauge field
\bea
\A_M &=& A_M + v_M \varphi
\eea
that is a supersymmetry invariant. If we define 
\bea
\D_M \Phi &=& \partial_M \Phi - i e [\A_M,\Phi]
\eea
then the standard gauge variations may be expressed in terms of $\A_M$ and $\D_M$ as  
\bea
\delta A_M &=& \D_M \lambda\cr
\delta \Phi &=& - i e [\Phi,\lambda]
\eea
So we see that $\A_M$ behaves just like a gauge field, with its corresponding field strength
\bea
\F_{MN} &=& F_{MN} + \D_M (v_N \varphi) - \D_N (v_M \varphi)
\eea
The vector multiplet Lagrangian can be recast in the following form
\bea
\L_{vector} &=& \frac{1}{2\lambda} \t{G}^{MN} \(\F_{MN} - 2 \D_M (v_N \varphi)\) + \L_{CS}(A) - \frac{1}{2} (\D_M \varphi)^2 - \frac{R}{10} \varphi^2\cr
&& + \frac{i}{2} \bar\psi \Gamma^M \D_M \psi 
\eea
with the supersymmetry variations 
\bea
\delta \varphi &=& - i \bar\eps \psi\cr
\delta \A_M &=& 0\cr
\delta \psi &=& \frac{1}{4\lambda} \Gamma^{MNP} \eps \(\F_{MN} - 2 \D_M (v_N \varphi)\) u_P - \frac{1}{\lambda} \Gamma^M \eps G_{MN} v^N + \Gamma^M \eps \D_M \varphi\cr
\delta G_{MN} &=& - \frac{i}{2} \D_Q \(\bar\eps \Gamma^Q \Gamma_{MNP} \psi\) u^P 
\eea
The Chern-Simons term can be expressed in terms of $\A_M - v_M \varphi$ in place of $A_M$ and then covariant derivatives with respect to $\A_M$ appear that act on $v_M \varphi$. All terms are quadratic in the fields, except for when $\D_M$ acts on some fields that gives in addition a factor of $\A_M$ that however is a supersymmetry single. What we have now achieved is that we have written the Lagrangian and the supersymmetry variations entirely in terms of the supersymmetry invariant gauge field $\A_M$ and there are no explicit commutator terms. The commutators that do appear are all implicit in the gauge covariant derivatives where the gauge field is a supersymmetry singlet. This means that we can now truncate supersymmetry in a different way from what we usually do under dimensional reduction, such that we can have a supersymmetric Lagrangian, even when $\L_v \eps$ is not vanishing \cite{Gustavsson:2021iex}. To this end we keep the zero modes of the bosonic fields. For the fermionic field, we keep the nonzero modes that correspond to the nonzero modes of the supersymmetry parameter $\eps$. To understand why this works, let us imagine that we have a six-dimensional action where we integrate the Lagrangian over all six directions, including the direction generated by $v^M$. Let us assume this direction is parametrized by a coordinate $\sigma \sim \sigma + 2\pi$. If we now keep the bosonic zero mode along $\sigma$, and if we make a supersymmetry variation of the Lagrangian, then the integral over $\sigma$ will select the modes of $\psi$ that match the modes of $\eps$ such that any fermionic bilinear combination $\bar\eps (...) \psi$ that appears in the variation of the action is a zero mode. For the supersymmetry variations of bosonic zero mode fields, we again integrate bilinear expressions $\bar\eps(...)\psi$ that selects the bosonic zero mode variation on the left-hand side, and those nonzero modes of $\psi$ that correspond to the modes of $\eps$ that makes the right-hand side a zero mode. For the supersymmetry variation of the fermionic field, the modes of the fermionic field will have to match with the mode of the supersymmetry parameter $\eps$ since these variations are on the form $\delta \psi = \eps \times$(bosonic zero mode field). We thus have a consistent truncation of the modes to the zero modes for the bosons and to the nonzero modes of the fermionic field that match the modes of the $\eps$. This is a consistent truncation of modes when the Lagrangian is abelian, or when the Lagrangian is nonabelian and all interactions come from covariant derivatives where the gauge field is a supersymmetry singlet. If there are other  interaction terms in the Lagrangian, then when we make a supersymmetry variation of the Lagrangian we may encounter the problematic term that is cubic in the fermionic field $\sim \bar\eps \psi \psi \psi$. For such cubic terms, integration over $\sigma$ will not just select one nonzero mode of the fermionic field, but it will select all modes in the entire Kaluza-Klein tower. In such cases there are no consistent truncations to finitely many modes, except for the trivial truncation down to both bosonic and fermionic zero modes, which is the usual dimensional reduction scenario. But if $\L_v \eps$ is non-zero, then the usual dimensional reduction of course breaks all supersymmetry. 

We now have got a supersymmetric Lagrangian, although just for the $(1,0)$ tensor multiplet, for any geometry restricted only by the existence of a solution to the six-dimensional conformal Killing spinor equation. But even then, we still have to restrict to the Lorentzian signature. It would therefore be interesting to see if lightlike dimensional reduction can be applied also in Euclidean signature. In Euclidean signature we can not impose the six-dimensional chiral Weyl projection since this is not compatible with the $SO(6,5)$ eleven-dimensional Majorana condition. So we are forced to consider the nonchiral $(2,2)$ tensor multiplet. Then we may obtain results in the $(2,0)$ theory by holomorphic factorization of  the $(2,2)$ theory. Lightlike dimensional reduction would in Euclidean signature correspond to dimensional reduction along a complex coordinate $\bar{z} = x - i t$, which means that the fields would depend holomorphically on $z$.

\section*{Acknowledgements}
This work was supported in part by NRF Grant RS-2023-00208011 and NRF Grant 2020R1I1A1A01052462.

\appendix
\section{Review of Sen's construction}\label{Sen}
Let us assume the gauge group is abelian. Then Lagrangian that was first proposed by Sen is given by\footnote{We do not consider tangent space selfduality for $Q_{MNP}$, so our approach coincides with Sen's approach only in flat Minkowski space.} \cite{Lambert:2019diy}, \cite{Andriolo:2020ykk}, \cite{Sen:2015nph}, \cite{Sen:2019qit}, \cite{Gustavsson:2020ugb}
\bea
\L_B &=& \frac{1}{24} H_{MNP}^2 + \frac{1}{6} Q^{MNP} H_{MNP}
\eea
where $Q^{MNP} = \frac{1}{6} \eps^{MNPRST} Q_{RST}$ is a selfdual Lagrange multiplier field that is imposing selfduality on $H_{MNP}$ as an equation of motion obtained by varying $Q^{MNP}$. The kinetic term for $H_{MNP}$ in this Lagrangian has the wrong sign. Nevertheless, we will take $H_{MNP}$ as the field strength of a two-form,
\bea
H_{MNP} &=& 3 \nabla_{[M} B_{NP]}
\eea
If we vary $B_{MN}$, then we get the equation of motion
\bea
\nabla_M \(h^{MNP} + 2 Q^{MNP}\) &=& 0
\eea
To understand the meaning of this equation of motion, we decompose $H_{MNP} = H_{MNP}^+ + H_{MNP}^-$ into selfdual and antiselfdual parts, and then we dualize
\bea
\eps^{MNPRST} \nabla_M \(H_{RST}^+ - H_{RST}^- + 2 Q_{MNP}\) &=& 0
\eea
Now we can compare this with the Bianchi identity
\bea
\eps^{MNPRST} \nabla_M \(H_{RST}^+ + H_{RST}^-\) &=& 0
\eea
By adding these two equations, we get
\bea
\nabla_{[M} \(Q_{RST]} + H_{RST]}^+\) &=& 0\label{Singlet}
\eea
that we solve by taking 
\bea
Q_{RST} &=& - H_{RST}^+ + C_{RST}^+
\eea
for some closed selfdual three-form $C_{MNP}^+$ satisfying
\bea
\nabla_{[M} C_{RST]}^+ &=& 0\label{wEOM}
\eea
Inserting this solution back into the Lagrangian, it becomes
\bea
\L_B &=& - \frac{1}{24} H_{MNP}^2 - \frac{1}{12} H^{MNP} C_{MNP}^+\label{wLag}
\eea
and now we get the right sign for the kinetic term.

Let us now study supersymmetry. We make the following ansatz for the supersymmetry variations,
\bea
\delta B_{MN} &=& i \bar\eps \Gamma_{MN} \psi\cr
\delta Q_{MNP} &=& - \frac{i}{2} \nabla_Q \(\bar\eps \Gamma^Q \Gamma_{MNP} \psi\)\label{QSUSY1}
\eea
where $\delta Q_{MNP}$ is selfdual as is necessary if we assume that $Q_{MNP}$ is selfdual offshell. The minus sign in the supersymmetry variation of $Q_{MNP}$ is chosen so that the equation (\ref{Singlet}) is invariant. The supersymmetry variation of the Lagrangian becomes
\bea
\delta \L_B &=& - \frac{i}{2} \bar\eps\Gamma_{NP}\psi \nabla_M Q^{MNP}
\eea
by using the Bianchi identity for $H_{MNP}$. For the matter part, we have the Lagrangian 
\bea
\L_m &=& - \frac{1}{2} (\nabla_M \phi^A) - \frac{R}{10} (\phi^A)^2 + \frac{i}{2} \bar\psi \Gamma^M \nabla_M \psi
\eea
Its supersymmetry variation becomes
\bea
\delta \L_m &=& - \frac{i}{12} \bar\psi \Gamma^Q \Gamma^{MNP} \eps \nabla_Q Q_{MNP}
\eea
by assuming the supersymmetry variation of the fermion is
\bea
\delta \psi &=& - \frac{1}{12} \Gamma^{MNP} \eps Q_{MNP} + {\mbox{matter}}\label{QSUSY2}
\eea
where we suppress the contributions from matter fields. Their contribution to $\delta \L_m$  cancel out among themselves in a standard manner. By using selfduality of $Q_{MNP}$ the variation can also be written as
\bea
\delta \L_m &=& \frac{i}{2}  \bar\eps\Gamma_{NP}\psi \nabla_M Q^{MNP}
\eea
and now we see that the full Lagrangian is supersymmetric, $\delta \L_H + \delta \L_m = 0$.

We may also work with the Lagrangian (\ref{wLag}). We shall then take $w^+_{MNP}$ to be a supersymmetry singlet. The supersymmetry variation then becomes
\bea
\delta \L_B &=& \frac{i}{4} \bar\eps\Gamma_{NP}\psi \nabla_M \(H^{MNP} + w^{+MNP}\)
\eea
For this to cancel against $\delta \L_m$, we shall change the supersymmetry variation to be
\bea
\delta \psi &=& \frac{1}{12} \Gamma^{MNP} \eps \(H_{MNP} + w_{MNP}^+\) + {\mbox{matter}}
\eea
We would now like to point out that there are minus signs here that are associated to $Q_{MNP}$, in equations (\ref{QSUSY1}) and (\ref{QSUSY2}) that are analogues to the minus signs that we found for $G_{ij}$ in this paper at those corresponding places.

\section{The supersymmetry variation of the Lagrangian}\label{Lm}
In this appendix we present the detailed computation for the supersymmetry variation of the  Lagrangian in the covariant formulation. For the matter part, 
\bea
\L_m &=& - \frac{1}{2} \(D_M \phi^A\)^2 - \frac{R}{10} (\phi^A)^2 \cr
&& + \frac{i}{2} \bar\psi \Gamma^M D_M \psi + \frac{e}{2} \bar\psi \Gamma_M \Gamma^A [\psi,\phi^A] v^M 
\eea
by omiting contributions from $G_{MN}$ and $F_{MN}$ in the variation of the fermion, we get
\bea
\delta \L_m &=& i \bar\eps \Gamma^A \psi D_M^2 \phi^A \label{1}\\
&& - \frac{i R}{5} \bar\eps \Gamma^A \psi \phi^A\label{2}\\
&& + i \bar\psi \Gamma^M \Gamma^N \Gamma^A \nabla_M \eps D_N \phi^A\label{3}\\
&& + i \bar\psi \Gamma^M \Gamma^N \Gamma^A \eps D_M D_N \phi^B\label{4}\\
&& - 4 i \bar\psi \Gamma^M \Gamma^A \nabla_M \eta \phi^A\label{45}\\
&& - 4 i \bar\psi \Gamma^M \Gamma^A \eta D_M \phi^A\label{5}\\
&& + \frac{e}{2} \bar\psi \Gamma^M \Gamma_N \Gamma^{AB} \nabla_M \eps [\phi^A,\phi^B] v^N\label{6}\\
&& + e \bar\psi \Gamma^M \Gamma_N \Gamma^{AB} \eps [D_M\phi^A,\phi^B] v^N\label{7}\\
&& + \frac{e}{2} \bar\psi \Gamma^{MN} \Gamma^{AB} [\phi^A,\phi^B] \nabla_M v_N\label{8}\\
&& - e \bar\psi \Gamma_M \Gamma^N \Gamma^A \Gamma^B \eps [D_N\phi^B,\phi^A]v^M\label{9}\\
&& - 4 e \bar\psi \Gamma_M \Gamma^{AB} \eta [\phi^B,\phi^A] v^M\label{10}\\
&& + \frac{i e^2}{2} \bar\psi \Gamma_M \Gamma_N \Gamma^A \Gamma^{BC} \eps [[\phi^B,\phi^C],\phi^A] v^P v^M\label{11}\\
&& - e D^M \phi^A [\bar\eps \Gamma_{MN} \psi,\phi^A] v^N\label{12}\cr
&& + \frac{ie}{2} \(\bar\psi \Gamma_N \Gamma^A [\psi,\bar\eps\Gamma^A\psi] - \bar\psi \Gamma^M [\psi,\bar\eps\Gamma_{MN} \psi]\) v^N\label{13}
\eea
We assume that $\eps$ is anticommuting. We expand (\ref{4}),
\bea
- i \bar\eps \Gamma^A \psi \phi^A + \frac{e}{2} \bar\psi \Gamma^{MN} \Gamma^A \eps [F_{MN},\phi^A]
\eea
and then the first term cancels (\ref{1}). We expand (\ref{3}),
\bea
- 4 i \bar\psi \Gamma^A \Gamma^N \eta D_N \phi^A
\eea
which cancels against (\ref{6}). On (\ref{5}) we apply $\Gamma^M \nabla_M \eta = - \frac{R}{20} \eps$, and get
\bea
- \frac{i R}{5} \bar\psi \Gamma^A \eta \phi^A
\eea
which cancels against (\ref{2}). We notice that (\ref{7}) and (\ref{11}) combine into 
\bea
2 e \bar\psi \Gamma^{AB} (v^M \nabla_M \eps) [\phi^A,\phi^B]
\eea
and this combines with (\ref{9}) into 
\bea
2 e \bar\psi \Gamma^{AB} (\L_v \eps) [\phi^A,\phi^B]
\eea
where the Lie derivative is given by
\bea
\L_v \eps &=& v^M \nabla_M \eps + \frac{1}{4} \nabla_M v_N \Gamma^{MN} \eps
\eea
We rewrite (\ref{8}) as
\bea
e \bar\psi \Gamma^{AB} \eps [\L_v \phi^A - i e [v^M A_M,\phi^A],\phi^B] + e \bar\psi \Gamma^{MN} \Gamma^{AB} \eps [D_M \phi^A,\phi^B] v_N
\eea
where the Lie derivative is 
\bea
\L_v \phi^A &=& v^M \nabla_M \phi^A
\eea
We rewrite (\ref{10}) as
\bea
&& - e \bar\psi \Gamma^{MN} \Gamma^{AB} \eps [D_M \phi^A,\phi^B] v_N - e \bar\psi \Gamma^{AB} \eps [v^M D_M \phi^A,\phi^B]\cr
&& - e \bar\psi \Gamma^{MN} \eps [D_M \phi^A,\phi^A] v_N - e \bar\psi \eps [v^M D_M\phi^A,\phi^A]
\eea
Then if we add (\ref{10}) and (\ref{8}) together, we get
\bea
&& 2 e \bar\psi \Gamma^{AB} \eps [\L_v \phi^A - i e [v^M A_M,\phi^A],\phi^B]  - e \bar\psi \eps [\L_v \phi^A - i [v^M A_M,\phi^A],\phi^A]\cr
&& + e \bar\psi \Gamma^{MN} \eps [D_M \phi^A,\phi^A] v_N\label{s}
\eea
We rewrite the last term as
\bea
e \bar\eps \Gamma^{MN} \psi [D_M \phi^A,\phi^A] v_N
\eea
and then we see that it cancels the last term in (\ref{s}). Line (\ref{13}) is identically zero by a Fierz identity \cite{Gustavsson:2020ugb}. Summing up what we have got,
\bea
\delta \L_m &=& 2 e \bar\psi \Gamma^{AB} (\L_v \eps) [\phi^A,\phi^B]\cr
&& + 2 e \bar\psi \Gamma^{AB} \eps [\L_v \phi^A - i e [v^M A_M,\phi^A],\phi^B] \cr
&&  - e \bar\psi \eps [\L_v \phi^A - i [v^M A_M,\phi^A],\phi^A]\cr
&& + \frac{e}{2} \bar\psi \Gamma^{MN} \Gamma^A \eps [F_{MN},\phi^A]\label{last}
\eea
To this we shall add the contributions from $F_{MN}$ and $G_{MN}$ that we omitted from $\delta \psi$, and then we shall add the supersymmetry variation of $\L_A$. 

We start with obtaining the supersymmetry variation of 
\bea
\L_A &=& \frac{1}{2\lambda} \t{G}^{MN} F_{MN} + \frac{1}{2\lambda} K^M K_M
\eea
The variation can be expressed as
\bea
\delta \L_A &=& \frac{1}{\lambda} \t{G}^{MN} D_M \delta A_N + \frac{1}{2\lambda} \delta G^{MN} \t{F}_{MN} + \frac{2}{\lambda} K^M u^N D_{[M} \delta A_{N]}
\eea
where in the second term, we have moved the tilde from $\t{G}^{MN}$ to $\t{F}_{MN}$, since it does not matter on which factor we remove the trace part. But at the level of computations, the difference $(\delta \t{G}^{MN}) F_{MN} - \delta G^{MN} \t{F}_{MN}$ turns out to be proportional to the Bianchi identity (\ref{Bianchi2}). We get 
\bea
\delta \L_A &=& - \delta A_N D_M \(\frac{1}{\lambda} \t{G}^{MN}\)\cr
&& + \frac{1}{2\lambda} \delta G^{MN} F_{MN}\cr
&& - \frac{1}{\lambda} \delta G^{MN} K_M v_N\cr
&& + \frac{1}{\lambda} K^M u^N D_M \delta A_N
\eea
where we can now apply the supersymmetry variation of $G^{MN}$. The explicit commutator term that is being produced by $K_M$ is now given by
\bea
\delta \L_A &=& - \frac{e}{2\lambda} \bar\eps \Gamma_Q \Gamma_{MNP} \Gamma^A \psi [K^M,\phi^A] u^N v^P v^Q
\eea
From $\delta \L_m$ we get a similar looking contribution by making the ansatz
\bea
\delta \psi &=& \Gamma^{MNP} \eps \alpha K_M u_N v_P
\eea
for some parameter $\alpha$. Then that term gives the commutator contribution
\bea
\delta \L_m &=& - e \alpha \bar\eps \Gamma_{MNP} \Gamma_Q \Gamma^A \psi [K^M,\phi^A]u^N v^P v^Q
\eea
The sum of these two contributions can be zero by choosing $\alpha = - \frac{1}{2\lambda}$ since then one finds the combination $\Gamma_{QMNP} v^P v^Q = 0$. We conclude that we should have 
\bea
\delta \psi &=& \Gamma^{MNP} \(\frac{1}{4\lambda} F_{MN} u_P - \frac{1}{4\lambda} \t{G}_{MN} v_P - \frac{1}{2\lambda} K_M u_N v_P\)
\eea
If we express this in terms of $\t{F}_{MN}$ it becomes
\bea
\delta \psi &=& \Gamma^{MNP} \(\frac{1}{4\lambda} \t{F}_{MN} u_P - \frac{1}{4\lambda} \t{G}_{MN} v_P - \frac{1}{\lambda} K_M u_N v_P\)
\eea

Let us next isolate all contributions involving $\t{G}^{MN}$. We have 
\bea
\delta \L_A &=& - i \bar\eps \Gamma_{NP} \psi v^P D_M \(\frac{1}{\lambda} \t{G}^{MN}\)
\eea
and 
\bea
\delta \L_m &=& i \bar\eps \Gamma_{NP} \psi v^P D_M \(\frac{1}{\lambda} \t{G}^{MN}\) + \frac{i}{2\lambda} \bar\eps \Gamma_{MN} \psi \L_v \t{G}_{MN}
\eea
so they cancel up to a Lie derivative.

Let us now list the remaining terms in $\delta \L_A$, 
\bea
\delta \L_A &=& \frac{i}{4} \bar\eps \Gamma^Q \Gamma^{MNP} \psi D_Q \(\frac{1}{\lambda} F_{MN} u_P + \frac{2}{\lambda} K_M u_N v_P\)\cr
&& - i \bar\eps \Gamma^{NP} \psi v_P D^M \(\frac{2}{\lambda} K_{[M} u_{N]}\)
\eea
From $\delta \L_m$ with the above $\delta \psi$, we get
\bea
\delta \L_m &=& - \frac{i}{4} \bar\eps \Gamma^{MNP} \Gamma^Q \psi D_Q \(\frac{1}{\lambda} F_{MN} u_P\) + \frac{i}{2} \bar\eps \Gamma^{MNP} \Gamma^Q D_Q \(\frac{1}{\lambda} K_M u_N v_P\)
\eea
Adding them, we get
\bea
\delta \L_A + \delta \L_m &=& \frac{i}{2} \bar\eps \Gamma^{QMNP} \psi F_{MN} D_Q\(\frac{1}{\lambda} u_P\)\cr
&& + \frac{i}{\lambda} \bar\eps \Gamma^{MN} \psi \L_v \(K_M u_N\)
\eea
We define
\bea
\Omega_{MN} &=& \nabla_M \(\frac{1}{\lambda} u_N\) - \nabla_N \(\frac{1}{\lambda} u_M\)
\eea
We can avoid the restriction to manifolds with $\Omega_{MN} = 0$ if we add a Chern-Simons term
\bea
\L_{CS} &=& - \frac{1}{4\lambda} \eps^{MNPQRS} \omega(A)_{MNP} \Omega_{QR} u_S
\eea
To see that, we notice the identity
\bea
\Omega_{MN} v^N &=& - \frac{1}{\lambda} \L_v u_M
\eea
so then we just need to assume that $\L_v u^M = 0$, which is a natural condition to impose on the manifold for the purpose of dimensional reduction, in order to have $\Omega_{MN} v^N = 0$. We choose the normalization such that $\delta \omega(A)_{MNP} = \delta A_M F_{NP}$. Then 
\bea
\delta \L_{CS} &=& - \frac{1}{4\lambda} \eps^{MNPQRS} i \bar\eps \Gamma_{MT} \psi F_{NP} \Omega_{QR} u_S v^T\cr
&=& - \frac{i}{4\lambda} \bar\eps \Gamma_{MT} \Gamma^{MNPQRS} \psi F_{NP} \Omega_{QR} u_S v^T\cr
&=& - \frac{i}{4\lambda} \bar\eps \Gamma^{NPQR} \psi F_{NP} \Omega_{QR} 
\eea
where we have used the identity
\bea
\Gamma_{MT} \Gamma^{MNPQRS}  &=& - 5 \delta_T^{[N} \Gamma^{PQRS]}
\eea
Adding up all the contributions, we get (\ref{deltaL}).

\end{document}